\documentclass[a4paper,12pt]{article}

\usepackage{jheppub} 
\usepackage{amsthm}
\usepackage{float}
\usepackage{braket}
\usepackage{latexsym}

\usepackage{gnuplot-lua-tikz}
\usepackage{ascmac}
\usepackage{qcircuit}
\usepackage{pxpgfmark}
\usepackage[T1]{fontenc}
\usepackage{lmodern}
\usepackage{tikz}
\usepackage{comment}
\usetikzlibrary{math, intersections,automata,positioning}

\theoremstyle{definition}
\newtheorem{thm}{Theorem}[section]
\newtheorem*{thm*}{Theorem}
\newtheorem{defn}[thm]{Definition}
\newtheorem*{defn*}{Definition}
\newtheorem{lem}[thm]{Lemma}
\newtheorem*{lem*}{Lemma}

\newtheorem*{rem*}{Remark}

\newtheorem*{con*}{Conjecture}

\newtheorem*{cor*}{Corollary}
\newtheorem{prop}[thm]{Proposition}
\newtheorem*{prop*}{Proposition}

\newtheorem*{hypoth*}{Hypothesis}

\newtheorem*{claim*}{Claim}
\newtheorem*{prf}{Proof}

\title{{Infinitely} Repeated Quantum Games and Strategic Efficiency}

\author[a]{Kazuki Ikeda}
\emailAdd{kazuki7131@gmail.com}
\author[b]{and Shoto Aoki}

\affiliation[a]{Department of Mathematics and Statistics
$\&$ Centre for Quantum Topology and Its Applications (quanTA), University of Saskatchewan, Saskatoon, Saskatchewan S7N 5E6, Canada}

\affiliation[b]{Department of Physics, Osaka University, Toyonaka, Osaka 5600043, Japan}

\date{\empty}
\arxivnumber{\empty}

\abstract{
Repeated quantum game theory addresses long term relations among players who choose quantum strategies. In the conventional quantum game theory, single round quantum games or at most finitely repeated games have been widely studied, however less is known for infinitely repeated quantum games. Investigating infinitely repeated games is crucial since finitely repeated games do not much differ from single round games. In this work we establish the concept of general repeated quantum games and show the Quantum Folk Theorem, which claims that by iterating a game one can find an equilibrium strategy of the game and receive reward that is not obtained by a Nash equilibrium of the corresponding single round quantum game. A significant difference between repeated quantum prisoner's dilemma and repeated classical prisoner's dilemma is that the classical Pareto optimal solution is not always an equilibrium of the repeated quantum game when entanglement is sufficiently strong. When entanglement is sufficiently strong and reward is small, mutual cooperation cannot be an equilibrium of the repeated quantum game. In addition we present several concrete equilibrium strategies of the repeated quantum prisoner's dilemma.
}

\begin{document} 
\maketitle 
\section{Introduction and Summary}
In our study on repeated quantum games, we aim at exploring long term relations among players and efficient quantum strategies that may make their payoff maximal and can become equilibria of the repetition of quantum games. The first work of an infinitely repeated quantum game was presented by one of the authors in \cite{2020QuIP...19...25I}, where infinitely repeated quantum games are explained in terms of quantum automata and optimal transformation and equilibria of some strategies are provided. The previous work is a quantum extension of a classical prisoner's dilemma where behavior is recognized by signals which can be monitored imperfectly or perfectly. To explore more on quantum properties of games, we give a different formulation of infinitely repeated games and investigate a long term relation between players. In this article we give a general definition of infinitely repeated quantum games. In particular we address an infinitely repeated quantum prisoner's dilemma and show the folk theorem of the repeated quantum game. The main contributions of our work to the quantum game theory are Theorem \ref{thm:anti} and Theorem \ref{thm:QFT}. By Theorem \ref{thm:anti}, we claim that mutual cooperation cannot be an equilibrium when players are maximally entangled and reward for cooperating is not enough. This fact distinguishes a repeated quantum game from a repeated classical game. Note that mutual cooperation is an equilibrium of the classical repeated prisoner's dilemma. By Theorem \ref{thm:QFT}, we present the quantum version of the Folk theorem, which assures the existence of equilibrium strategies for any degree of entanglement. {Our model will also be useful in discussing complex dynamic processes and evolutionary processes in many economic or interdisciplinary fields from a game theoretical perspective. In economics, the negotiation of a contract can be regarded as a repeated game if it continues for a long time, and the recent Internet auction is basically a repeated game. Most recently, a quantum model of contracts, moral hazard, and incentive contracts was first formulated in~\cite{Ikeda_cat2021}. It would be very interesting to consider such a model in terms of repeated games. Furthermore, considering the evolutionary nature of ecological systems will be a new challenge~\cite{doi:10.1098/rspa.2021.0397}}. For further motivation to consider repeated quantum games, please consult \cite{2020QuIP...19...25I, 2020arXiv201014098A} for example. 
 
In the sense of classical games, players try to maximize reward by choosing strategies based on opponents' signals obtained by measurement. In a single stage prisoner's dilemma, mutual cooperation is not a Nash equilibrium \cite{Nash48} but a Pareto optimal, however when repetition of the game is allowed, such a Pareto optimal strategy can be an equilibrium of the infinitely repeated game. A study on an infinitely repeated game aims at finding such a non-trivial equilibrium solution that can be established in a long term relationship. From this viewpoint repeated games play a fundamental role in a fairly large part of the modern economics. Hence repeated quantum games will become important when quantum systems become prevalent in many parts of the near future society.    

There are various definitions of quantum game \cite{Eisert:1998vv,1999PhRvL..82.1052M} and it is as open to establish the foundation. Indeed, many authors work on quantum games with various motivations \cite{Li2013,Li13}. Recent advances in quantum games are reviewed in \cite{2018arXiv180307919S}. Most of the conventional works on quantum games focus on some single stage quantum games, where each one plays a quantum strategy only one time. However, in a practical situation, a game is more likely played repeatedly, hence it is more meaningful to address repeated quantum games. Repeated games are categorized into (1) finitely repeated games and (2) infinitely repeated games. However, equilibria of finitely repeated classical games are completely the same as those of single round classical games. This is also true for at most finitely repeated quantum games and the Nash equilibrium of finite repeated quantum prisoner's dilemma is the same as the single stage case \cite{2002PhLA..300..541I,2012JPhA...45h5307F}.  

On the other hand, in infinitely repeated (classical) games or unknown length games, there is no fixed optimum strategy, hence they are very different from single stage games \cite{10.2307/1911077,10.2307/1911307}. It is widely known that in infinitely repeated (classical) prisoner's dilemma, mutual cooperation can be an equilibrium. So it is natural to investigate similar strategic behavior of quantum players. But less has been known for infinitely repeated quantum games. To our best knowledge, this is the first article which gives a formal definition of infinitely repeated games and investigates the general Folk theorem of the infinitely repeated prisoner's dilemma. 

This work is organized as follows. In Sec.\ref{sec:SRQG}, we address single round quantum prisoner's dilemma and investigate equilibria of the game. We first prepare terminologies and concepts used for this work. We present novel relations between payoff and entanglement in Sec.\ref{sec:equi}. In Sec.\ref{sec:RQG} we establish the concept of a generic repeated quantum game (Definition \ref{def:RQG}) and present some equilibrium strategies (Lemma \ref{lem:1} and \ref{lem:2}). Our main results (\ref{thm:anti} and \ref{thm:QFT}) are described in Sec.\ref{sec:Folk}. This work is concluded in Sec.\ref{sec:fin}.

\section{Single Round Quantum Games}
\label{sec:SRQG}
\subsection{Setup}
We consider the quantum prisoner's dilemma~\cite{Eisert:1998vv,1999PhRvL..82.1052M}. Let $\ket{C},\ket{D}$ be two normalized orthogonal states that represent "Cooperative" and "Non-cooperative" states, respectively. Then a general state of the game is a complex linear combination, spanned by the basis vectors $\ket{CC}, \ket{CD}, \ket{DC}, \ket{DD}$. Each player $i$ chooses a unitary strategy $U_i$ and the game is in a state
\begin{align}
\Ket{\psi}=\mathcal{J}^\dagger (U_A\otimes U_B) \mathcal{J}\Ket{CC}, 
\end{align}
where $\mathcal{J}$ gives some entanglement. In what follows we use 
\begin{align}
    \mathcal{J}=\exp{\left[i\frac{\theta}{2}Y\otimes Y\right]}
    =\cos{\frac{\theta}{2} }+i\sin{\frac{\theta}{2}}(Y\otimes Y), 
\end{align}
where $\theta$ represents entanglement and two states become maximally entangled at $\theta=\frac{\pi}{2}$. We assume a quantum strategy of a player $i$ has a representation
\begin{align}
\label{eq:U}
    U_i=\alpha_iI +i \left(\beta_i X+ \gamma_i Y + \delta_i Z\right) \quad (\alpha_i, \cdots ,\delta_i \in \mathbb{R})
\end{align}
where the coefficients obey $\alpha^2_i+\cdots +\delta_i^2=1$. 

\begin{table}[H]
    \centering
\begin{tabular}[t]{c|cc|}
 & Bob:C & Bob:D \\ \hline
Alice:C & ($r$,$r$) & ($s$,$t$) \\
Alice:D & ($t$,$s$) & ($p$,$p$) \\ \hline
\end{tabular}
    \caption{Pay-off matrix of the prisoner's dilemma. $p,r,s$ and $t$ satisfy $t>r>p>s\ge0$. }
    \label{fig:payoff}
\end{table}

Alice receives payoff $\$_A=\Braket{\psi|\hat{\$}_A|\psi}$, where $\hat{\$}_A$ is her payoff matrix
\begin{align}
    \hat{\$}_A=r\Ket{CC}\bra{CC}+p\ket{DD}\bra{DD}+t\ket{DC}\bra{DC}+s\ket{CD}\bra{CD}. 
\end{align}
Without loss of generality, {we can always proceed with the discussion as $s=0$ if necessary}. Taking into account of $U_A$ \eqref{eq:U}, the average payoff $\hat{\$}_A=\hat{\$}_A(x_A,x_B)$ of Alice is a function of $x_i=(\alpha_i, \cdots ,\delta_i)$
\begin{align}
\begin{split}
    \$_A(x_A,x_B)&=r(\alpha_A \alpha_B-\delta_A\delta_B+\sin{\theta}(\beta_A\gamma_B+\gamma_A\beta_B))^2 \\
    &+p(\beta_A\beta_B-\gamma_A\gamma_B+\sin{\theta}(\alpha_A\delta_B+\delta_A\alpha_B))^2 \\
    &+t(\gamma_A\alpha_B+\beta_A\delta_B+\sin{\theta}(-\alpha_A\beta_B+\delta_A\gamma_B))^2 \\
    &+\cos^2{\theta}\left(r(\alpha_A\delta_B+\delta_A\delta_B)^2+p(\beta_A\gamma_B+\gamma_A\beta_B)^2 \right. \\
    &\left.+t(\beta_A\alpha_B-\gamma_A\delta_B)^2 \right). 
\end{split} \label{form6}
\end{align}
Since the game is symmetric for two players, $\$_A(x_A,x_B)=\$_B(x_B,x_A)$ is satisfied. 


\begin{defn}
A quantum strategy $U$ is called pure if a player chooses any single qubit operator.
\end{defn}
\begin{defn}
\label{def:mixed}
A quantum strategy $U$ is called mixed if a player choose one from $\{I,X,Y,Z\}$ with some probability $(p_I,p_X,p_Y,p_Z)$\footnote{The most general definition of a single-qubit mixed strategy $U$ is given as
\begin{equation}
    U=\int_{U(2)} p(g)g d\mu(g), 
\end{equation}
where $d\mu$ is a Haar measure on $U(2)$ and $p(g)$ is non-negative and satisfies $\int_{U(2)} p(g)d\mu(g)=1$. Throughout this article, we intend to explore mixed strategies based on Def. \ref{def:mixed}. 
}.
\end{defn}

Although our mixed strategies are probabilistic mixtures of the four Pauli operators $\{I,X,Y,Z\}$, we show the game is powerful enough to obtain stronger results compared with classical prisoner’s dilemma with mixed strategy. 

The classical pure strategic prisoner's dilemma allows players to choose one of $\{I,X\}$ and it becomes a mixed strategy game if the state is written as 
\begin{equation}
    \Ket{\psi}=\sum_{i\in\{CC,CD,DC,DD\}}p_i\ket{i}, 
\end{equation}
where $p_i$ is non-negative and satisfies $\sum_ip_i=1$. Note that for a single round quantum game, a generic state is represented as 
\begin{equation}
    \ket{\psi}=\sum_{i\in\{CC,CD,DC,DD\}}c_i\ket{i},
\end{equation}
where $c_i$ is a complex number and a state $\ket{i}$ is measured with the probability $|c_i|^2$. However as long as one considers a single round game, the quantum game is resemble the classical mixed strategy game, where a strategy is chosen stochastically with a given probability distribution. 

Considering repeated quantum games is a way to make quantum games much more meaningful, since transition of quantum states is different from that of classical states. To begin with, we first address single round quantum prisoner's dilemma and equilibria of the game. 

\subsection{Equilibria}
\label{sec:equi}
\subsubsection{Without Entanglement $\theta=0$}
The case without entanglement corresponds to the classical prisoner's dilemma. Table \ref{table:t2} presents the correspondence between operators and classical actions. 

\begin{table}[H]
    \centering
\begin{tabular}[t]{c|cccc|}
 & I & X & Y & Z \\ \hline
I & (C,C) & (C,D) & (C,D) & (C,C) \\
X & (D,C) & (D,D) & (D,D) & (D,C) \\
Y & (D,C) & (D,D) & (D,D) & (D,C) \\
Z & (C,C) & (C,D) & (C,D) & (C,C) \\ \hline
\end{tabular}
    \caption{Operators and actions for the case without entanglement $\theta=0$.}
    \label{table:t2}
\end{table}

Then the strategy is simply a mixed strategy of the pure strategies $\{C,D\}$, therefore average payoff of Alice is  
\begin{align}
\begin{split}
    \$_A&=r(\alpha_A^2+\delta_A^2)(\alpha_B^2+\delta_B^2) \\
    &+p(\beta_A^2+\gamma_A^2)(\beta_B^2+\gamma_B^2) \\
    &+t(\beta_A^2+\gamma_A^2)(\alpha_B^2+\delta_B^2). 
\end{split}
\end{align}
As a result, $U_A=i(\beta_A X +\gamma_A Y), U_B=i(\beta_B X +\gamma_B Y)$ is the equilibrium. 

\subsubsection{Maximally Entangled $\theta=\pi/2$}
In contrast to the case without entanglement, the strategy part of the maximally entangled case becomes  
\begin{align}
    \mathcal{J}^\dagger( I\otimes X )\mathcal{J}&=(I\otimes X)\mathcal{J}^2=-Y\otimes Z \\
    \mathcal{J}^\dagger( I\otimes Y )\mathcal{J}&=I\otimes Y. 
\end{align}
Therefore the correspondence between the operators and actions becomes non-trivial as exhibited in Table \ref{table:t3}. 

\begin{table}[H]
    \centering
    \begin{tabular}[t]{c|cccc|}
 & I & X & Y & Z \\ \hline
I & (C,C) & (D,C) & (C,D) & (D,D) \\
X & (C,D) & (D,D) & (C,C) & (D,C) \\
Y & (D,C) & (C,C) & (D,D) & (C,D) \\
Z & (D,D) & (C,D) & (D,C) & (C,C) \\ \hline
\end{tabular}
    \caption{Operators and actions for the maximal entangled case $\theta=\frac{\pi}{2}$.}
    \label{table:t3}
\end{table}

\paragraph{Pure Quantum Strategy}
We first consider the pure strategy game. The average payoff of Alice is 
\begin{align}
\begin{split}
    \$_A(x_A,x_B)&=r(\alpha_A \alpha_B-\delta_A\delta_B+\beta_A\gamma_B+\gamma_A\beta_B)^2 \\
    &+p(\beta_A\beta_B-\gamma_A\gamma_B+\alpha_A\delta_B+\delta_A\alpha_B)^2 \\
    &+t(\gamma_A\alpha_B+\beta_A\delta_B-\alpha_A\beta_B+\delta_A\gamma_B)^2 \label{form5}.
\end{split}
\end{align}
This can be written as 
\begin{align}
    \$_A=r(\alpha_A^\prime)^2+p(\beta_A^\prime)^2+t(\gamma_A^\prime)^2,
\end{align}
where 
\begin{align}
\label{eq:form3}
\begin{split}
\alpha_A^\prime&=\alpha_A \alpha_B-\delta_A\delta_B+\beta_A\gamma_B+\gamma_A\beta_B \\
    \beta_A^\prime&=\beta_A\beta_B-\gamma_A\gamma_B+\alpha_A\delta_B+\delta_A\alpha_B \\
    \gamma_A^\prime&=\gamma_A\alpha_B+\beta_A\delta_B-\alpha_A\beta_B+\delta_A\gamma_B \\
    \delta_A^\prime&=\gamma_B\alpha_A+\beta_B\delta_A-\alpha_B\beta_A+\delta_B\gamma_A
\end{split}
\end{align}
Note that the equation \eqref{eq:form3} is a coordinate transformation on $S^3$, hence there is a $x_A\in S^3$ that makes $\gamma_A^\prime=1$ for all $x_B\in S^3$. In other words, Alice can find a stronger strategy for any strategy of Bob, and vice versa. Therefore there is no equilibrium for this case. In this sense this looks like "Rock-paper-scissors" .

\paragraph{Mixed Quantum Strategy}
We can find equilibria when we allow Alice and Bob play mixed quantum strategies. That means Alice chooses a strategy from $\{I,X,Y,Z\}$ with probability $p^A=(p^A_I,p^A_X,p^A_Y,p^A_Z)$. Alice can execute her strategy $U_A$ in such a way that 
\begin{align}
\label{eq:MS}
    U_A=\sqrt{p_I^A}I\otimes I+\sqrt{p_X^A} iX \otimes I+\sqrt{p_Y^A} iY \otimes X +\sqrt{p_Z^A} iZ \otimes X 
\end{align}

Then the average payoff of Alice is given by 
\begin{align}
    \$_A=\left(
    \begin{array}{cccc}
        p^A_I & p^A_X & p^A_Y & p^A_Z 
    \end{array} \right) \left(
    \begin{array}{cccc}
        r & t & s & p \\  
        s & p & r & t \\
        t & r & p & s \\
        p & s & t & r 
    \end{array} \right) \left(
    \begin{array}{c}
         p^B_I  \\
         p^B_X \\
         p^B_Y \\
         p^B_Z
    \end{array} \right) \label{form4}
\end{align}
We find that the above form can be written as 
\begin{align}
        \$_A&=r p^B_I + t p^B_X + s p^B_Y +p p^B_Z \nonumber \\
        &+\left(
        \begin{array}{ccc}
             p^A_X & p^A_Y & p^A_Z 
        \end{array} \right) \left(
        \begin{array}{c}
             (r-s)(-p^B_I+p^B_Y)+(t-p)(-p^B_X+p^B_Z)  \\
             (t-r)(p^B_I-p^B_X)+(p-s)(p^B_Y-p^B_Z) \\
             (r-p)(-p^B_I+p^B_Z)+(t-s)(p^B_Y-p^B_X) 
        \end{array} \right) 
\label{eq:form8}
\end{align}

If $p^B_I=p^B_X=p^B_Y=p^B_Z=\frac{1}{4}$, the average payoff of Alice does not depend on $p^A$ and it becomes 
\begin{align}
    \$_A=\frac{r+t+s+p}{4}. 
\end{align}
So one of the equilibria is 
\begin{align}
\left\{
\begin{array}{c}
     {p^A}^\star=\left(\frac{1}{4},\frac{1}{4},\frac{1}{4},\frac{1}{4}\right) \\
    {p^B}^\star=\left(\frac{1}{4},\frac{1}{4},\frac{1}{4},\frac{1}{4}\right)
\end{array}\right.
\end{align}

In addition, we have another equilibrium. 
\begin{prop}
If $t+s>r+p$, then 
\begin{align}
\left\{
\begin{array}{c}
     {p^A}^\star=\left(0,\frac{1}{2},\frac{1}{2},0\right) \\
    {p^B}^\star=\left(\frac{1}{2},0,0,\frac{1}{2}\right)
\end{array}\right.
~~\text{or}~~
\left\{
\begin{array}{c}
     {p^A}^\star=\left(\frac{1}{2},0,0,\frac{1}{2}\right) \\
    {p^B}^\star=\left(0,\frac{1}{2},\frac{1}{2},0\right)
\end{array}\right.
\label{eq:form9}
\end{align}
is an equilibrium. The average payoff at the equilibrium is $\frac{t+s}{2}$. 
\end{prop}
\begin{prf}
It is enough to show one of \eqref{eq:form9}. Using \eqref{eq:form8}, we find that the average payoff of Alice respects 
\begin{align}
    \$_A(p^A,{p^B}^\star)&=\frac{r+p}{2}+(p^A_X+p^A_Y)\frac{1}{2}(t-p-r+s) \\
    &\leq \frac{t+s}{2}=\$_A({p^A}^\star,{p^B}^\star)
\end{align}
Similarly the average payoff of Bob satisfies 
\begin{align}
    \$_B({p^A}^\star,p^B)&=\frac{t+s}{2}-(p^B_X+p^B_Y)\frac{1}{2}(t-p-r+s) \\
    &\leq \frac{t+s}{2}=\$_B({p^A}^\star,{p^B}^\star). 
\end{align}
This ends the proof. 
\qed 
\end{prf}
   
Repeating the same discussion, we can show the following statement.
\begin{prop}
If $t+s<r+p$, then 
\begin{align}
\left\{
\begin{array}{c}
     {p^A}^\star=\left(0,\frac{1}{2},\frac{1}{2},0\right) \\
    {p^B}^\star=\left(0,\frac{1}{2},\frac{1}{2},0\right)
\end{array}\right.
~~\text{or}~~
\left\{
\begin{array}{c}
     {p^A}^\star=\left(\frac{1}{2},0,0,\frac{1}{2}\right) \\
    {p^B}^\star=\left(\frac{1}{2},0,0,\frac{1}{2}\right)
\end{array}\right.
\label{eq:form10}
\end{align}
is an equilibrium. The average payoff at the equilibrium is $\frac{r+p}{2}$. 
\end{prop}

\subsubsection{In-between}
\paragraph{Pure Quantum Strategy}
 We consider a general case with entanglement $\theta\in(0,\pi/2)$. Without entanglement $\theta=0$, both $X,Y$ are always stronger than $I$, but $I$ is sometimes stronger than $X$ when there is some entanglement. $Y$ is always stronger than $I$. So in what follows we first assume Bob plays $U_B=Y$, which corresponds to $x_B=(0,0,1,0)$. In this case, the payoff of Alice is 
\begin{align}
\begin{split}
 \$_A(x_A,x_B)&=(r\sin^2\theta +p\cos^2\theta) \nonumber \\
    &-(p\cos^2\theta-(t-r)\sin^2\theta)\delta_A^2 \nonumber\\
    &-(r-p)\sin^2\theta\gamma_A^2 \nonumber \\
    &-(r\sin^2\theta +p\cos^2\theta)\alpha_A^2. 
\end{split}
\end{align}
Note that the first term is always positive, the third and fourth terms are always negative since $r>p>0$. The second term could be positive if $\sin^2\theta <\frac{p}{t-r+p}$. Therefore Alice's best reaction to $U_B=Y$ is $x_A=(\alpha_A,\beta_A,\gamma_A,\delta_A)=(0,1,0,0)$ and the corresponding payoff is 
\begin{equation}
    \$_A(x_A^\star,x_B^\star)=r\sin^2\theta +p\cos^2\theta.
\end{equation}

Similarly one can find that Bob's best reaction to $x_A=(0,1,0,0)$ is $x_B=(0,0,1,0)$. So the pair of $x^\star_A=(0,1,0,0)$ and $x^\star_B=(0,0,1,0)$ is an equilibrium since they satisfy
\begin{align}
\begin{split}
     \$_A(x_A^\star,x_B^\star)\geq \$_A(x_A,x_B^\star) \\
    \$_B(x_A^\star,x_B^\star)\geq \$_A(x_A^\star,x_B)
\end{split}
\end{align}

More generally one can find that 
\begin{align}
\left\{
\begin{array}{l}
    x_A^\star=(0,\cos\phi,\sin\phi,0)\\
    x_B^\star=(0,\sin \phi,\cos \phi,0) 
\end{array} \right. \label{form7}
\end{align}
is an equilibrium for all $\phi \in [0,2\pi]$ and the corresponding average payoff of Alice is \begin{equation}
    \$_A(x_A^\star,x_B^\star)=r\sin^2\theta +p \cos^2\theta. 
\end{equation}  

Equilibrium strategies do not exist if $\sin^2\theta >\frac{p}{t-r+p}$, which is consistent with our previous discussion that the maximal entanglement $\theta=\frac{\pi}{2}$ case does not have any equilibrium while only pure quantum strategies are played. 

\paragraph{Mixed Quantum Strategy}
In general the average payoff of Alice can be written as
\begin{align}
    \$_A&=\left(
    \begin{array}{cccc}
        p^A_I & p^A_X & p^A_Y & p^A_Z 
    \end{array} \right)  A\left(
    \begin{array}{c}
         p^B_I  \\
         p^B_X \\
         p^B_Y \\
         p^B_Z
    \end{array} \right) \label{form11}\\ 
    A&=\left(
    \begin{array}{cccc}
        r & s\cos^2\theta+t \sin^2\theta & s & r\cos^2\theta+ p\sin^2\theta \\  
        t\cos^2\theta +s \sin^2\theta  & p & p\cos^2\theta +r\sin^2\theta & t \\
        t & p\cos^2\theta+r\sin^2\theta & p & t\cos^2\theta+s\sin^2\theta \\
        r\cos^2\theta+p\sin^2\theta & s & s\cos^2\theta+t\sin^2\theta & r 
    \end{array} \right) 
\label{eq:payoff2}
\end{align}

We define 
\begin{align}
    \begin{array}{c}
         \Bar{t}=t\cos^2\theta +s \sin^2\theta \\
         \Bar{r}=r\cos^2\theta +p \sin^2\theta \\
         \Bar{p}=p\cos^2\theta +r \sin^2\theta \\
         \Bar{s}=s\cos^2\theta +t \sin^2\theta
    \end{array} 
\end{align}
and 
\begin{align}
\begin{split}
        p_I^{\star}=p_Z^{\star}=\frac{1}{2}\frac{s+\Bar{s}-p-\Bar{p}}{t+\Bar{t}-r-\Bar{r}-p-\Bar{p}+s+\Bar{s}} \\
    p_X^{\star}=p_Y^{\star}=\frac{1}{2}\frac{t+\Bar{t}-r-\Bar{r}}{t+\Bar{t}-r-\Bar{r}-p-\Bar{p}+s+\Bar{s}}
\end{split}
\end{align}

\begin{prop}
\label{prop:1}
$p^A=p^B=p^\star$ is an equilibrium 
\end{prop}
\begin{prf}
The equation \eqref{eq:payoff2} can be decomposed into 
\begin{align}
    \$_A&=\left(
    \begin{array}{cccc}
        r & \Bar{s} & s & \Bar{r} 
    \end{array} \right)  \left(
    \begin{array}{c}
         p^B_I  \\
         p^B_X \\
         p^B_Y \\
         p^B_Z
    \end{array} \right) \nonumber \\
    &+\left(
    \begin{array}{ccc}
        p_X^A & p_Y^A & p_Z^A  
    \end{array} \right) \left(
    \begin{array}{cccc}
         \Bar{t}-r & p-\Bar{s} & \Bar{p}-s & t-\Bar{r} \\
         t-r & \Bar{p}-\Bar{s} & p-s & \Bar{t}-\Bar{r} \\
         \Bar{r}-r & s-\Bar{s} & \Bar{s}-s & r-\Bar{r}
    \end{array} \right)\left(
    \begin{array}{c}
         p^B_I  \\
         p^B_X \\
         p^B_Y \\
         p^B_Z
    \end{array} \right) 
\end{align}
One can show that the second term vanishes at $p^B=p^\star$. Therefore $\$_A(p^A,p^\star)=\$_A(p^\star,p^\star)$ is satisfied for any $p^A$. Since the same argument is held for Bob's average payoff, $(p^A,p^B)=(p^\star,p^\star)$ is an equilibrium.   \qed 
\end{prf}

In addition, we can find another equilibrium for $t+s>r+p$ and $t+s<r+p$. We first address the case where $t+s>r+p$. 

\begin{prop}
\label{prop:2}
If $t+s>r+p$, the following pairs of $(p^A,p^B)$ are equilibria. 
\begin{align}
    \left\{
\begin{array}{c}
     {p^A}^\star=\left(0,\frac{1}{2},\frac{1}{2},0\right) \\
     {p^B}^\star=\left(0,\frac{1}{2},\frac{1}{2},0\right)
\end{array}\right. \quad \left(\sin^2\theta<\frac{2(p-s)}{t-r+p-s} \right) \\
    \left\{
\begin{array}{c}
    {p^A}^\star=\left(\frac{1}{2},0,0,\frac{1}{2}\right) \\
     {p^B}^\star=\left(0,\frac{1}{2},\frac{1}{2},0\right)
\end{array}\right. \quad \left(\sin^2\theta>\frac{2(p-s)}{t-r+p-s} \right) 
\end{align}
\end{prop}

\begin{prf}
We first consider the case where $\sin^2\theta<\frac{2(p-s)}{t-r+p-s}$ holds. We write ${p^A}^\star={p^B}^\star=\left(0,\frac{1}{2},\frac{1}{2},0\right)$. Then Alice's average payoff is 
\begin{align}
    \$_A(p^A,{p^B}^\star)&=
    \frac{1}{2}\left(s\cos^2\theta + t\sin^2\theta+s \right)\nonumber \\
    &+\frac{1}{2} (
    -(t-r)\sin^2\theta +(p-s)(\cos^2\theta+1))
    ({p^A_X+p^A_Y})\label{eq:form12}
\end{align}
Note that the second term is positive for $\left(\sin^2\theta<\frac{2(p-s)}{t-r+p-s} \right)$. Therefore $\$_A({p^A}^\star,{p^B}^\star)\ge \$_A(p^A,{p^B}^\star)$ for all $p^A$. In the same way, the average payoff of Bob also respects $\$_B({p^A}^\star,{p^B}^\star)\ge \$_A({p^A}^\star,{p^B})$. 

For $\sin^2\theta>\frac{2(p-s)}{t-r+p-s}$, we write ${p^A}^\star=\left(\frac{1}{2},0,0,\frac{1}{2}\right),{p^B}^\star=\left(0,\frac{1}{2},\frac{1}{2},0\right)$. Then the average payoff of Alice is exactly the same as \eqref{eq:form12}. Since $\sin^2\theta>\frac{2(p-s)}{t-r+p-s}$, the second term of \eqref{eq:form12} is negative. Hence it becomes maximal for $p^A_X=p^A_Y=0$, namely $\$_A({p^A}^\star,{p^B}^\star)\ge \$_A(p^A,{p^B}^\star)$ for all $p^A$. The average payoff of Bob is 
\begin{align}
    {\$_B}({p^A}^\star,{p^B})&=
    \frac{1}{2}\left(r\cos^2\theta + p\sin^2\theta+r \right)\nonumber \\
    &+\frac{1}{2} (
    -(p-s)\sin^2\theta +(t-r)(\cos^2\theta+1)
    ) (p^B_X+p^B_Y)
\end{align}
Since $t+s>r+p$, the second term is always positive. Therefore $\$_B({p^A}^\star,{p^B}^\star)\ge \$_B({p^A}^\star,p^B)$ for all $p^B$. This ends the proof. 
\qed 
\end{prf}

Note that for $\sin^2\theta>\frac{2(p-s)}{t-r+p-s}$ and at the equilibrium, the average payoff of Alice is smaller than that of Bob: $\$_A({p^A}^\star,{p^B}^\star)<\$_B({p^A}^\star,{p^B}^\star)$.

\begin{figure}[H]
    \centering
    \includegraphics[width=11cm]{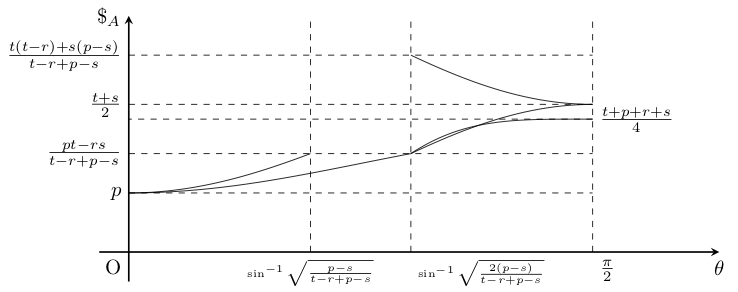}
    \caption{The average payoff of Alice at the Nash equilibrium, when $t+s>r+p$.}
    \label{fig:graph1}
\end{figure}

In summary, the average payoff of Alice is exhibited in Fig.\ref{fig:graph1}.  

We next consider the case where $t+s<r+p$. One can show the following statement as we did before.  
\begin{prop}
\label{prop:3}
If $t+s<r+p$, the following pairs of $(p^A,p^B)$ are equilibria. 
\begin{align}
    &\left\{
\begin{array}{c}
     {p^A}^\star=\left(0,\frac{1}{2},\frac{1}{2},0\right) \\
    {p^B}^\star=\left(0,\frac{1}{2},\frac{1}{2},0\right) 
\end{array}\right. \quad \left(0\leq \theta \leq \frac{\pi}{2} \right) \label{eq:form13}\\ 
    &\left\{
\begin{array}{c}
    {p^A}^\star=\left(\frac{1}{2},0,0,\frac{1}{2}\right) \\
     {p^B}^\star=\left(\frac{1}{2},0,0\frac{1}{2}\right)
\end{array}\right. \quad \left(\sin^2\theta>\frac{2(t-r)}{t-r+p-s} \right)  \label{eq:form14}
\end{align}
\end{prop}

\begin{figure}[H]
    \centering
    \includegraphics[width=11cm]{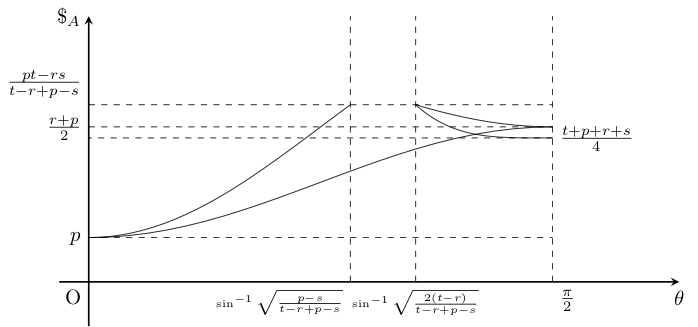}
    \caption{The average payoff of Alice at the Nash equilibrium, when $t+s<r+p$ and $2(t-r)>p-s$.}
    \label{fig:graph2}
\end{figure}

\begin{figure}[H]
    \centering
    \includegraphics[width=11cm]{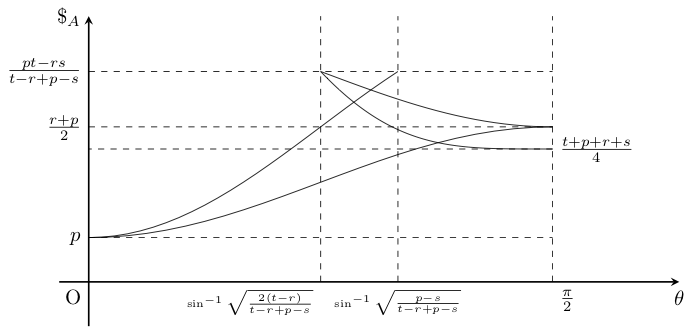}
    \caption{The average of Alice's payoff at the Nash equilibrium, when $t+s<r+p$ and $2(t-r)<p-s$.}
    \label{fig:graph3}
\end{figure}

Since $1<\frac{2(p-s)}{t-r+p-s}$ is always true, they satisfy $\sin^2\theta<\frac{2(p-s)}{t-r+p-s}$. Therefore \eqref{eq:form13} is an equilibrium for all $\theta$. In summary, the average payoff of Alice is exhibited in Fig.\ref{fig:graph2} and \ref{fig:graph3}.  

\section{Repeated Quantum Games}
\label{sec:RQG}
\subsection{Setup}
A generic repeated quantum game is summarized as follows. 
\begin{defn}[Repeated Quantum Game]
\label{def:RQG}
Let $\{\tau_i\}_{i=1,2,\cdots}$ be a set of positive integers and $\{\ket{\psi^t}\}_{t=0,1,\cdots}$ be a family of quantum states. For each $i$, put $t_i=\sum_k^i\tau_k$. Let $\{\ket{i}\}_i$ be an orthonormal basis and $\{c_i\}_i$ be a set of complex numbers such that $\sum_i|c_i|^2=1$. With respect to each $\ket{i}$, suppose that reward to a player $n$ is given by a certain operator $\hat{\$}_n$ in such a way that $\bra{i}\hat{\$}_n\ket{i}$ and $\bra{i}\hat{\$}_n\ket{j}=0$ for $i\neq j$. Then a general $N$-persons quantum repeated game proceeds as follows:
\begin{enumerate}
    \item At any $t$, the game is in a state 
    \begin{equation}
        \ket{\psi^t}=\sum_{i}c_i\ket{i}, 
    \end{equation}
    \item The time-evolution of the game is given by unitary operators $U^t$
    \begin{equation}
        \ket{\psi^{t+1}}=U^t\ket{\psi^t}
    \end{equation}
    \item At time $t=t_i$, reward is evaluated and given to players $n$ by a 
    \begin{equation}
        \$^{t_i}_n=f(\tau_i)\bra{\psi^{t_i}}\hat{\$}_n\ket{\psi^{t_i}},
    \end{equation}
    where $f(\tau_i)$ is a function such that $f(1)=1$. {It is a collective expression of the interest and other costs incurred during that period of $\tau_i$.}  
    \item The total payoff of a player $n$ at time $t\in[t_k, t_{k+1})$ is defined by 
    \begin{equation}
        \$_n(t)=\sum_i^{k} \delta^{t_i}\$^{t_i}_n
    \end{equation}
    and in the limit it is $\$_n=\lim_{t\to\infty}\$_n(t)$. 
\end{enumerate}
\end{defn}

Now let us consider the repeated quantum prisoner's dilemma. Let $\ket{\psi^t}\in \mathcal{H}_A \otimes \mathcal{H}_B$ be a state at the $t$-th round. Let $\{\tau_i\}_{i=1,2,\cdots}$ be a set of positive integers. We assume measurement of quantum states is done at the end of the $t_i=\sum_{k=1}^i\tau_k$ round for each $i$. For example, $\tau_i=1$ for all $i$ means we measure quantum states every round. Suppose $V^{\tau_{i-1}+1}, \cdots, V^{\tau_{i}}$ are operators of Alice, then her strategy is $U^{t_i}_A=V^{t_{i-1}+1}_A\otimes\cdots \otimes V^{t_{i-1}+\tau_{i}}_A$ and the game is in a state  
\begin{equation}
    \ket{\psi^{t_i}}=\mathcal{J}^\dagger(U^{t_i}_A\otimes U^{t_i}_B)\mathcal{J}\Ket{\psi^{t_{i-1}}}
\end{equation}
which should be measured at the end of $t_i$ round. The payoff could depend on $\tau_i$:
\begin{equation}
    \$^{t_i}_A=f(\tau_{i})\bra{\psi^{t_i}}\hat{\$}_A\Ket{\psi^{t_i}},
\end{equation}
where $f(\tau_i)$ is a certain function of $\tau_i$. In repeated games, the total payoff is written with discount factor $\delta \in(0,1)$ and in our case it can be introduced in such a way that 
\begin{equation}
    \$_A=\sum_{i}\delta^{t_i}\$^{t_i}_A.
\end{equation}
An interpretation of the discount factor is that the importance of the future payoffs decreases with time. The measurement periods $\{\tau_i\}$ can be defined randomly or predeterminedly. When game's state is measured every round, the probability of monitoring signals is classical (see \eqref{eq:MS}). In order to enjoy full quantum games, a game should evolve with unitary operation. 

Since this work is the first study on infinitely repeated quantum games, we focus on the most fundamental case and investigate a role of entanglement. We will work on a generic case elsewhere. For this purpose, in what follows, we put $\tau_i=1$ for all $i$ and $f(1)=1$. {Currently, we are considering only the prisoner's dilemma, but there are frequent cases when the game is repeated, for example in negotiation games. Negotiations usually involve thinking not only about the pie on the table today, but also about the pie after tomorrow. It is natural that interest may accrue and costs may change during negotiations. The function $f$ represents the change in cost that can occur during the period between the finalization of one state and the transition to the next state. Negotiation games are the basic repeated games in contract theory. The formulation of contract theory as quantum games has been recently given in~\cite{Ikeda_cat2021}. Further developing that model as a game theory is a very interesting problem.}

Then the game evolves in such a way that 
\begin{align}
\Ket{\psi^{t+1}}=\mathcal{J}^\dagger (U_A^t\otimes U_B^t) \mathcal{J}\Ket{\psi^t}, 
\end{align}
where $U_A^t$ is her quantum strategy for the $t$th round. Then her average payoff is 
\begin{equation}
    \$^t_A=\Braket{\psi^t|\hat{\$}_A|\psi^t}. 
\end{equation}

\subsection{Equilibria}
We define a trigger strategy as follows. 
\begin{defn}[Trigger 1]
\begin{enumerate}
    \item Alice and Bob play $I$ at time $\tau$ if cooperative relation is maintained before time $\tau$. 
    \item If Alice (Bob) deviates from this cooperative strategy at $\tau$, then Bob (Alice) chooses either $X$ or $Y$ with equal probability at time $\tau+1$. 
\end{enumerate}
\end{defn}

We show that Trigger 1 is an equilibrium of the repeated quantum prisoner's dilemma. 
\begin{lem}
\label{lem:1}
Trigger 1 is an equilibrium of the repeated quantum prisoner's dilemma if either one of the following conditions are satisfied:
\begin{enumerate}
    \item $r+p>t+s$
    \item $r+p<t+s$ and $\sin^2\theta<\frac{2(p-s)}{t-r+p-s}$
\end{enumerate}
. 
\end{lem} 
\begin{prf}
Suppose they play $I$ for all $t$. Then Alice's total payoff is 
\begin{align}
\begin{split}
     V_A&=r+\delta r+\delta^2 r+\cdots\\
     &=\frac{r}{1-\delta}
\end{split}
\end{align}
Now assume that they play $I$ before $t$ and Alice plays $Y$ at $\tau$. Since Bob plays $I$ at $\tau$, her maximal payoff is 
\begin{align}
    \begin{split}
        V'_A&=t+\delta\$_A(p^A, {p^B}^\star)+\delta^2\$_A(p^A, {p^B}^\star)+\cdots\\
        &=t+\frac{\delta}{1-\delta}\$_A(p^A, {p^B}^\star), 
    \end{split}
\end{align}
where ${p^B}^\star=\left(0,\frac{1}{2},\frac{1}{2},0\right)$. Note that, according to Propositions \ref{prop:2} and \ref{prop:3}, ${p^B}^\star$ is an equilibrium of the game. 

Choosing $Y$ cannot be her incentive if $V_A>V'_A$, which is equilibrium to 
\begin{equation}
    r-t+\frac{\delta}{1-\delta}(r-\$_A(p^A,{p^B}^\star))>0. 
\end{equation}
Since $r-\$_A(p^A,{p^B}^\star)$ is always positive, this inequality is satisfied for a large $\delta$. The greatest lower bound $\delta_{\inf}$ of such $\delta$ is obtained as  
\begin{equation}
    \delta_{\inf}=\frac{t-r}{t-\frac{1}{2}(p+p\cos^2\theta+r\sin^2\theta))}
\end{equation}
Since $\delta_{\inf}$ is smaller than 1, no deviation occurs. 

If Alice or Bob deviates from cooperative strategy, they choose $X$ or $Y$ with equal probability along trigger 1. According to Proposition \ref{prop:2} and \ref{prop:3}, choosing $X$ or $Y$ with equal probability each other is an equilibrium when condition 1,2 are satisfied. So no deviation occurs, as long as Alice and Bob choose $X$ or $Y$ with equal probability. 

Therefore, trigger 1 can be an equilibrium of the repeated game.
\qed 
\end{prf}

\begin{figure}
    \centering
    \includegraphics[width=11cm]{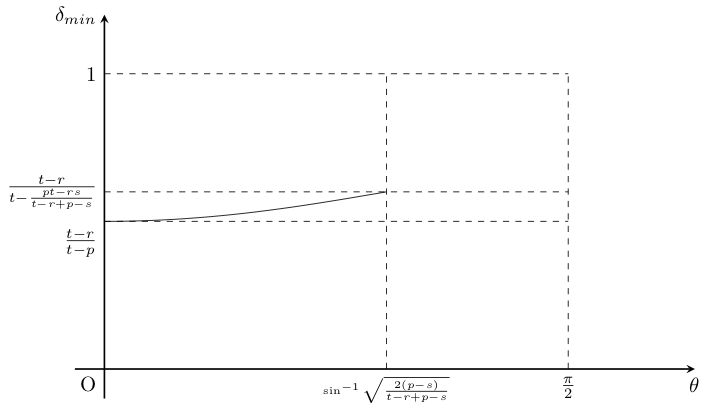}
    \caption{The greatest lower bound $\delta_{\inf}$ of discount factor as a function of the entangle parameter $\theta$ when $t+s>r+p$.}
    \label{fig:graph4}
\end{figure}

\begin{figure}
    \centering
    \includegraphics[width=11cm]{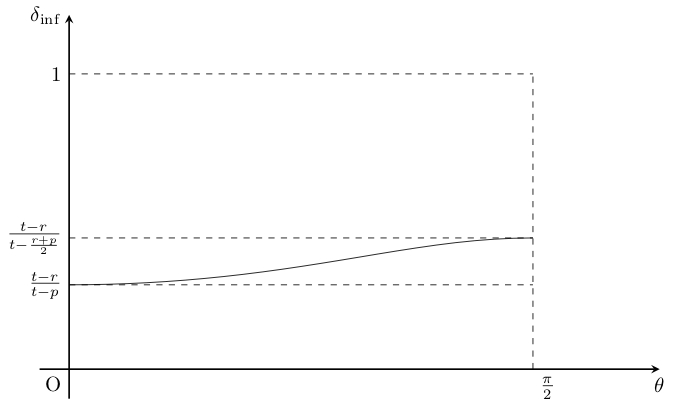}
    \caption{The greatest lower bound $\delta_{\inf}$ as a function of the entangle parameter $\theta$ when $t+s<r+p$.}
    \label{fig:graph5}
\end{figure}

Fig.\ref{fig:graph4} and \ref{fig:graph5} present the relation between $\delta_{\inf}$ and entanglement. The bigger $\theta$ becomes, the more profit the players can receive, therefore the discount factor $\delta$ should be sufficiently big in order to maintain the cooperative relation when their states are strongly entangled. 

\begin{defn}[Trigger 2]
\begin{enumerate}
    \item Alice and Bob play $I$ at time $\tau$ if cooperative relation is maintained before time $\tau$. 
    \item If Bob (Alice) deviates from the cooperative strategy at time $\tau$, then Alice (Bob) chooses either $X$ or $Y$ with the equal probability at time $\tau+1$.
    \item If Alice and Bob plays $X$ and $Y$ (or $Y$ and $X$) respectively at time $\tau$, then at time $\tau+1$ Alice (Bob) repeats the same strategy, otherwise chooses either $X$ or $Y$ with the equal probability at time $\tau+1$.
    \item If Alice and Bob repeat $X$ and $Y$ (or $Y$ and $X$) respectively before time $\tau$ and Bob (Alice) deviates from the repetition at time $\tau$, then Alice (Bob) chooses either $X$ or $Y$ with the equal probability at time $\tau+1$.
\end{enumerate}
\end{defn}

\begin{figure}
    \centering
    \includegraphics[width=12cm]{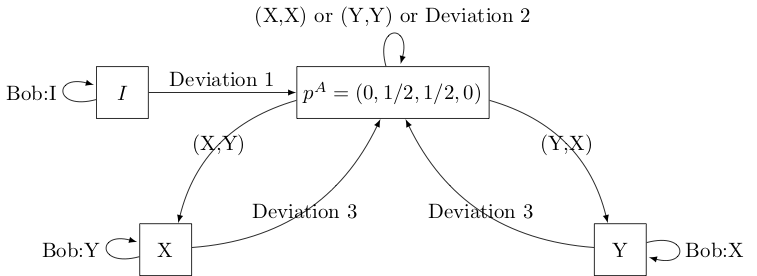} 
    \caption{Alice's strategy diagram against Bob's strategies}
    \label{fig:auto2}
\end{figure}
 
Deviation 1 means that Alice (Bob) deviates the cooperation and chooses $X$ or $Y$. Deviation 2 means that Alice (Bob) chooses $I$ or $Z$ when they should choose $X$ or $Y$ with the equal probability. According to Proposition \ref{prop:2}, it can happen when $\sin^2\theta > \frac{2(p-s)}{t-r+p-s}$. Deviation 3 means that Alice (Bob) chooses $I$ or $Z$ when Alice should play $X$ ($Y$) and Bob should play $Y$ $(X)$, respectively. It can happen at $\sin^2\theta > \frac{p-s}{t-r+p-s}$. This is because that if an opponent chooses $Y$ $(X)$, choosing $Z$ $(I)$ against $X$ $(Y)$ can yield more profit.

We can modify Trigger 1 and redefine Trigger 2 so that for all $\theta$, it can be a subgame perfect equilibrium. 

\begin{lem}[Trigger 2]
\label{lem:2}
Trigger 2 is an equilibrium of the repeated game for $r>\frac{t+p}{2}$. 
\end{lem}
\begin{prf}
We first address the case where Alice choose deviation 1. Her total expected payoff when she plays $I$ is $V_A=\frac{r}{1-\delta}$. If Alice chosen deviation 1, Alice and Bob choose $X$ or $Y$ with the equal probability. $(X,X), (X,Y), (Y,X), (Y,Y)$ can occur with the equal probability. In this round, the expected payoff is $\frac{1}{2}\left(p+p\cos^2\theta +r\sin^2\theta\right)$. If $(X,X)$ or $(Y,Y)$ is played, they should choose $X$ or $Y$ in the next round. If $(X,Y)$ or $(Y,X)$ is played, they should choose the same quantum strategy as last time in the next round. Therefore her maximal expected payoff for choosing deviation 1 is 
\begin{align}
\begin{split}
    V'_A&=t+\sum_{t=1}^{+\infty}\delta^t\left(\frac{1}{2} \right)^t\left( \$_A(X,X)+\$_A(X,Y) \right)+\sum_{s=1}^{+\infty}\sum_{t=1}^{+\infty}\delta^{s+t}\left( \frac{1}{2} \right)^s\$_A(X,Y) \\
    &=t+\frac{\frac{\delta}{2}}{1-\frac{\delta}{2}}\left( p+\frac{1}{1-\delta}\left(p\cos^2\theta+r\sin^2\theta \right)\right)
\end{split}    
\end{align}
She does not have an incentive to choose deviation 1 when $V_A-V'_A>0$, which can happen is $\delta$ is large.  

Regarding deviation 2, Alice's total expected payoff when she choose either $X$ or $Y$ with the equal probability is 
\begin{equation}
    V_A=\frac{\frac{1}{2}}{1-\frac{\delta}{2}}\left( p+\frac{1}{1-\delta}\left(p\cos^2\theta+r\sin^2\theta\right)\right)
\end{equation}
and her maximal expected payoff for choosing deviation 2 is 
\begin{equation}
    V'_A=\frac{s\cos^2\theta+t\sin^2\theta}{2}+\delta V_A. 
\end{equation}
She does not have an incentive to choose deviation 2 if $V_A-V'_A>0$, which is satisfied for a large $\delta$.  

On deviation 3, Alice's total expected payoff for choosing $X$ ($Y$) while Bob choosing $Y$ ($X$) is 
\begin{align}
\begin{split}
     V_A&=\frac{\$_A(X,Y)}{1-\delta}\\
     &=\frac{p\cos^2\theta+r\sin^2\theta}{1-\delta}
\end{split}
\end{align}
Her maximal expected payoff for choosing deviation 3 is 
\begin{equation}
    V'_A=s\cos^2\theta+t\sin^2\theta+\delta\left( \frac{\frac{1}{2}}{1-\frac{\delta}{2}}\left( p+\frac{1}{1-\delta}\left(p\cos^2\theta+r\sin^2\theta\right)\right)\right)
\end{equation}
since Alice and Bob choose $X$ or $Y$ with the equal probability along  trigger 2. She does not have an incentive to choose deviation 3 if $V_A-V'_A>0$, which is satisfied for a large $\delta$. 

For all cases, one can find $\delta<1$ that does not endows a player with an incentive for deviation. 
\qed 
\end{prf}

\begin{figure}[H]
    \centering
    \includegraphics[width=11cm]{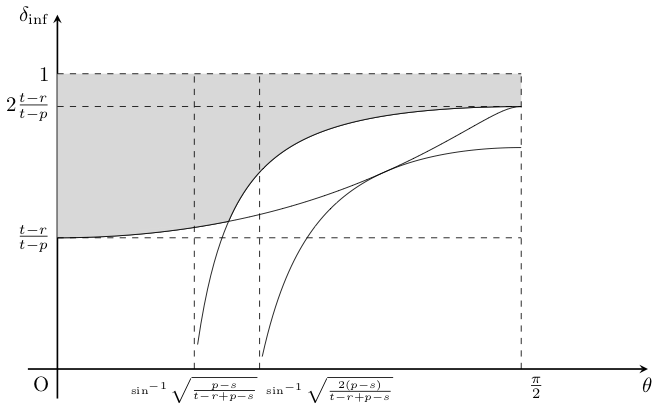}
    \caption{The region {(gray)} where trigger 2 becomes an equilibrium when $r>\frac{t+p}{2}$.}
    \label{fig:graph6_2}
\end{figure}

The curves in Fig. \ref{fig:graph6_2} presents the relation between $\theta$ and the greatest lower bound $\delta_{\inf}$ of $\delta$ that respects the condition $V_A-V'_A>0$. Such $\delta$ lives in the colored region in the figure. 

\subsection{Folk Theorem}
\label{sec:Folk}
\subsubsection{Pure Quantum Strategy}
\begin{figure}[H]
    \centering
    \includegraphics[width=8cm]{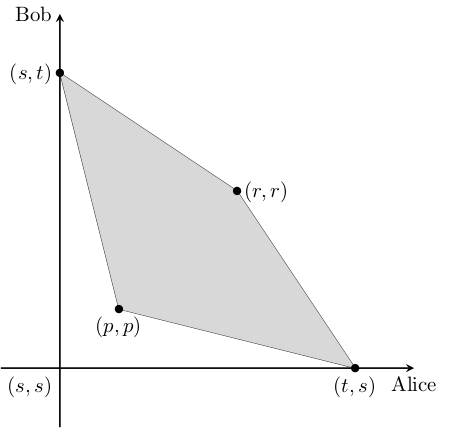}
    \caption{Domain of profit for $r>\frac{t+s}{2}$.}
    \label{fig:folk1}
\end{figure}

\begin{figure}[H]
    \centering
    \includegraphics[width=8cm]{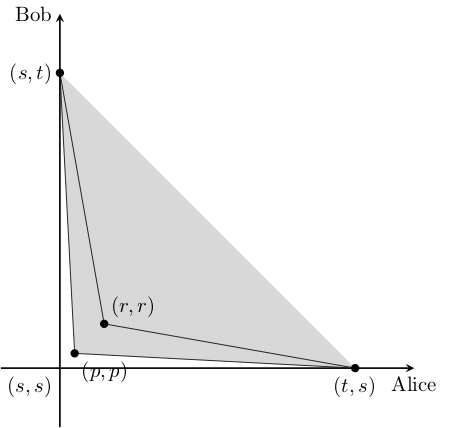}
    \caption{Domain of profit for $r<\frac{t+s}{2}$.}
    \label{fig:folk2}
\end{figure}

We first consider both players choose pure strategies. Let $S=U(2)\times U(2)$ be a set of strategies of Alice and Bob. We write a pair of Alice and Bob's profit $\$(a)=\left(\$_A(a),\$_B(a) \right)$ for a strategy $a\in S$. And let $\nu$ be the profit per single round of the repeated game. Then it turns out that they can receive profit $\nu$ in $V$ (Fig.\ref{fig:folk1} and \ref{fig:folk2}).  

\begin{align}
    V=\left\{\nu \in \mathbb{R}^2 \middle| \nu =\int_{ S}p(a)\$(a) d\mu(a) ,p(a)\geq0,\int_{S} p(a)d\mu(a) =1\right\},
\end{align}
where $d\mu$ is a Haar measure on $U(2)\times U(2)$. 

\begin{figure}[H]
    \centering
    \includegraphics[width=11cm]{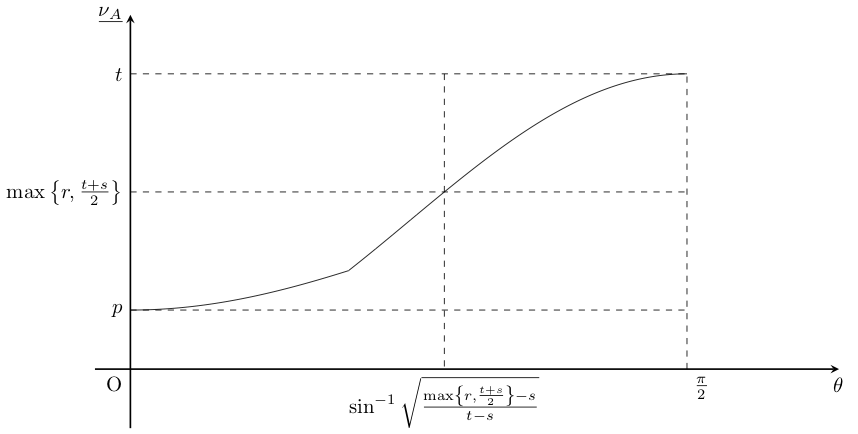}
    \caption{The mini-max value of Alice when both Alice and Bob choose only pure strategies.}
    \label{fig:graph7_1}
\end{figure}

Alice's mini-max value is defined as 
\begin{align}
    \underline{\nu_A}=\inf_{U_B} \sup_{U_A}\$_A(U_A,U_B), 
\end{align}
which is shown in Fig.\ref{fig:graph7_1}. If $\underline{\nu_A}$ is smaller than $\max\left\{r,\frac{t+s}{2}\right\}$, then there exists a subgame perfect equilibrium. This can be summarized as follows. 

\begin{thm}
While players choose pure quantum strategies, there exists a subgame perfect equilibrium of the repeated quantum prisoner's dilemma if 
\begin{align}
    \sin^2 \theta <\frac{\max\left\{r,\frac{t+s}{2}\right\}-s}{t-s}. 
\end{align}
\end{thm}

\subsubsection{Mixed Quantum Strategy}
\begin{figure}[H]
    \centering
    \includegraphics[width=11cm]{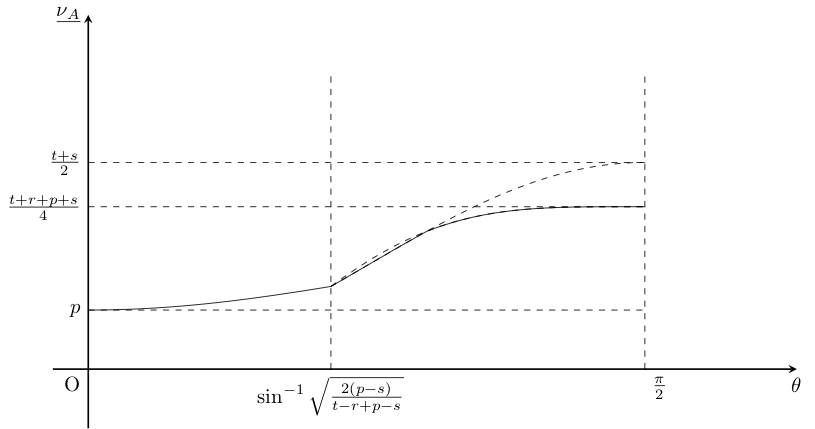}
    \caption{The mini-max value of Alice for $t+s>r+p$ when they play mixed strategies.}
    \label{fig:graph7_2}
\end{figure}

\begin{figure}[H]
    \centering
    \includegraphics[width=11cm]{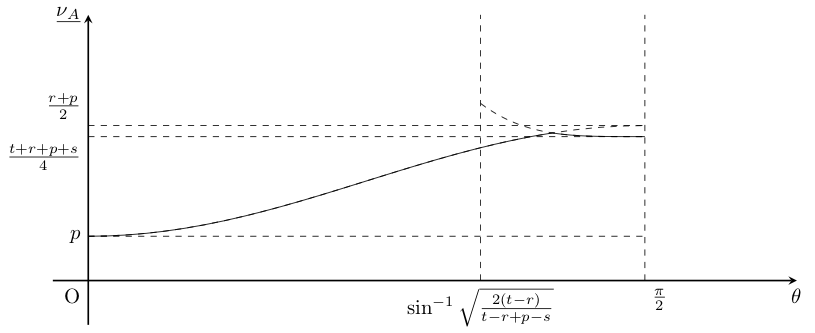}
    \caption{The mini-max value of Alice for $t+s<r+p$ when they play mixed strategies.}
    \label{fig:graph7_3}
\end{figure}

\begin{figure}[H]
    \centering
    \includegraphics[width=8cm]{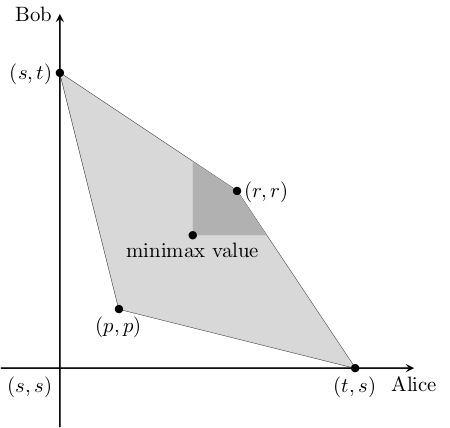}
    \caption{The feasible and individually rational payoff set for $r>\frac{t+s}{2}$. {Since the dark gray area contains $(r,r)$, in this case the solution in which both cooperate is realized as the quantum version of the equilibrium solution (Quantum Folk Theorem).}}
    \label{fig:folk1_2}
\end{figure}

\begin{figure}[H]
    \centering
    \includegraphics[width=8cm]{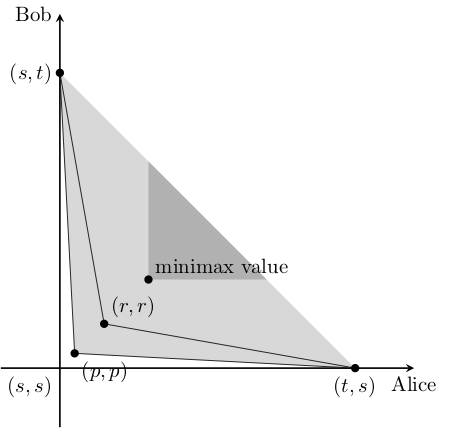}
    \caption{The feasible and individually rational payoff set for $r<\frac{t+s}{2}$. {Since the dark gray area does not contain $(r,r)$, in this case the solution in which both cooperate is not realized as the quantum version of the equilibrium solution (Anti-Folk Theorem).}}
    \label{fig:folk2_2}
\end{figure}

The mini-max value for the mixed strategy case is defined as
\begin{align}
    \underline{\nu_A}=\min_{{p^B}}\max_{{p^A}}{\$_A}({p^A},{p^B}),
\end{align}
which is shown in Fig\ref{fig:graph7_2} and \ref{fig:graph7_3}. There is a subgame perfect equilibrium since $\underline{\nu_A}<\max\left\{r,\frac{t+s}{2}\right\}$ is always satisfied. Note that, for all entanglement parameter $\theta$ there exists non-empty region 
\begin{align}
    V^\star=\left\{\nu \in V \middle| \nu_i>\underline{\nu_i},i=A,B \right\},
\end{align}
which we call the feasible and individually rational payoff set. If Alice and Bob are entangled, the mini-max value becomes large and the area of $V^\star$ gets small. Especially, the area of $V^\star$ becomes the smallest when states are maximally entangled $\theta=\frac{\pi}{2}$ and is exhibited as the dark parts of Fig.\ref{fig:folk1_2} and \ref{fig:folk2_2}. It does not contain $(r,r)$ when $\theta=\frac{\pi}{2}$, therefore a notable difference between classical and quantum repeated prisoner's dilemma is the following. 

\begin{thm}[Anti-Folk Theorem]
\label{thm:anti}
When Alice and Bob are maximally entangled, cooperation and cooperation cannot be realized unless $r\ge\frac{t+s}{2}$.   
\end{thm}

In the classical repeated prisoner's game, cooperation and cooperation is the Pareto optimal solution and can be an equilibrium of the repeated game, called the Folk theorem, whereas it is not true for the quantum repeated game. For the case of our quantum game, one can always find a better solution that makes their payoff larger than choosing cooperation and cooperation. So in order to establish the cooperative relation, they require a sufficiently large reward $r$ for cooperating. 

Based on a similar argument for the classical Folk theorem, we obtain its quantum version. 

\begin{thm}[Quantum Folk Theorem]
\label{thm:QFT}
For all entanglement parameter $\theta$, there exists an appropriate discount factor $\delta\in(0,1)$ such that any payoff in the feasible and individually rational payoff are in the set $V^\star$. 
\end{thm}

\begin{prf}
Let $m_A$ be Alice's mixed strategy that endows Bob with the mini-max value.  Then if the round-average payoff for playing $a^\star \in A$ is in $V^\star$, the following is an equilibrium. 
\begin{enumerate}
    \item Choose $a^\star\in A$ unless deviation occurs.
    \item If deviation is observed, the both play $m=(m_A,m_B)$ for the next $T$ rounds, then the both play $a^\star$.
    \item Once if $m$ is not chosen in the $T$ rounds, the both again choose $m$ for the succeeding $T$ rounds.     
\end{enumerate}
Here $T$ is chosen as 
\begin{align}
    d_i<T(\$_i(a^\star)-\$_i(m)), \quad (i=A,B)
\end{align}
where $d_i=\max_{a_i} \{\$_i(a_i,a_{-i}^\star)-\$_i(a^\star)\}$. 

One can complete the proof of the above as did in \cite{10.2307/1911307}. 
\qed 
\end{prf}

\section{Conclusion and Future Directions}
\label{sec:fin}
In this work we established the concept of a generic repeated quantum game and addressed repeated quantum prisoner's dilemma. Based on Definition \ref{def:RQG}, we addressed the repeated quantum prisoner's dilemma for the case of $\tau_i=1$ for all $i$. Even though the game evolves in a classical manner, repetition of the quantum game is much different from the classical repeated prisoner's dilemma (Theorem \ref{thm:anti}) and obtained a novel feasible and individually rational set $V^\star$ for the repeated quantum prisoner's dilemma (Theorem \ref{thm:QFT}). 

{As stated at the beginning of this article, repeated games play a very important role in game theory. In many cases, human activities are followed by long-term strategic relationships. In particular, in order to form a sustainable market, it is necessary to consider the interests and strategies of market participants and develop rules from the perspective of repeated games. It is a market on complex Hilbert space, which is not envisioned by conventional economic theory, but will be realized in the near future with the realization of the quantum Internet. As this paper has shown, the equilibrium solution of a classical game is not necessarily the solution in an infinite quantum game. This clearly shows that infinite iteration quantum games are not an extension of classical theory, and suggests that quantum economics on complex Hilbert spaces is completely different from conventional economic theory. From this point of view, it is clear that the economics of quantum money~\cite{2020arXiv201014098A}, market design and contract theory with quantum operations~\cite{Ikeda_cat2021} need to be seriously investigated, and that these are new and challenging issues in the quantum age.}

There are many research directions of the repeated quantum games. For example, it will be interesting to address cases where $\tau_i\neq 1$. In this work we focused on the $\tau_i=1$ case and as a result we clarified a role of entanglement $\theta$. From a viewpoint of quantum computation, the complex probability coefficients $c_i$ are important as well as entanglement. Taking into account of it, studying more on a generic time-evolution of the quantum game is crucial and it is necessary to introduce general review periods $\tau_i$ which the players should not know a priori.

Furthermore, it is important to address games with three or more persons from a viewpoint of repeated quantum games. As Morgenstern and Neumann described \cite{morgenstern1953theory}, games with three or more persons can be much different from games with two persons, since they have a chance to form a coalition. In quantum games, there are various ways to introduce entanglement which makes games more complicated. In the future quantum network, people will use entangled strategies in general for various purposes. Therefore it will be more practical to investigate quantum games where many strategies are entangled. More recently the authors investigated a quantum economy where three persons create, consume and exchange some entangled quantum information, using entangled strategies \cite{2020arXiv201014098A}. According to it, quantum people can use stronger strategies than people using only classical resources, and their strategic behavior in a quantum market can be much more complicated in general. So as emphasized in \cite{2020arXiv201014098A}, economic in the quantum era will become essentially different from that today in the literature. From this perspective, investigating a long term relation among quantum players will be significant.

Moreover it is also interesting to apply quantum repeated games not only to economics but also to other fields, since game theoretic perspectives are used in various disciplines such as business administration, psychology, biology, and computer science. Studying quantum effects of strategies and human behavior in a global quantum system will require advanced quantum technologies, but quantum games will also shed new light on a study completed in a local quantum system. For example, recent advances in applications of quantum algorithm are crucial for potential speedup of machine learning, which can be regarded as a repeated game. Generative adversarial networks (GANs) could be such a typical example. In general, GANs have multiple and complicated equilibria and it will be interesting to explore QGANs \cite{PhysRevLett.121.040502} in terms of repeated quantum games. 

 We leave it an open question to investigate a Nash equilibrium in general repeated quantum games. For single stage non-cooperative games, the equilibria in pure quantum strategies are addressed in \cite{2018arXiv180102053S} and those in mixed quantum strategies are given in \cite{1999PhRvL..82.1052M}, by using Glicksburg's fixed-point theorem.

\section*{Acknowledgement}
This work was supported by PIMS Postdoctoral Fellowship Award (K. I.)


\bibliographystyle{JHEP}
\bibliography{ref}
\end{document}